\documentclass[cameraready]{Interspeech}
\usepackage{xcolor}
\usepackage[normalem]{ulem}
\definecolor{deepBlue}{HTML}{0D47A1}

\title{IndicContextEval: A Benchmark for Evaluating Context Utilisation in Audio Large Language Models Across 8 Indic Languages}

\author[affiliation={1}, equalcontribution]{Sakshi}{Joshi}

\author[affiliation={2}, equalcontribution]{Dhruv Subhash}{Rathi}

\author[affiliation={2}, equalcontribution]{Sanskar}{Singh}

\author[affiliation={1}]{Eldho Ittan}{George}
\author[affiliation={1}]{R J}{Hari}

\author[affiliation={1}]{Kaushal}{Bhogale}

\author[affiliation={1}]{Mitesh M.}{Khapra}

\address{
    $^1$AI4Bharat,Indian Institute of Technology Madras, India\quad$^2$Sarvam AI, India
    }

\email{sakshijcom@gmail.com, miteshk@dsai.iitm.ac.in}

\newcommand{\benchmark}{IndicContextEval}

\usepackage{comment}
\usepackage{multirow}
\usepackage{booktabs}
\usepackage{url}
\usepackage{pgfplots}
\pgfplotsset{compat=1.18}
\usepackage{colortbl}
\begin{document}

\maketitle

\begin{abstract}
AudioLLMs enable speech recognition conditioned on textual prompts such as domain descriptions or entity lists. However, it remains unclear whether these models genuinely utilise such context or rely on parametric knowledge learned during pretraining. Existing benchmarks cannot answer this question because they evaluate transcription under fixed prompting conditions and rarely include explicit contextual inputs. We introduce \benchmark, a 56-hour multilingual benchmark of natural speech from 555 speakers across 8 Indian languages and 23 professional domains. We design a 7-level prompting framework that progressively introduces contextual signals, including metadata, natural-language descriptions, entity lists in English and native script, and adversarial prompts with incorrect entities. Evaluating five models reveals substantial differences in context utilisation behaviour, highlighting the need for explicit evaluation of contextual grounding in AudioLLMs.
\end{abstract}
\section{Introduction}

Automatic speech recognition systems are increasingly deployed in applications where contextual information is available at inference time. For example, meeting transcription systems may know the meeting topic, medical dictation systems have access to domain terminology, and voice assistants often maintain user-specific entity lists. Such context can help resolve ambiguities in the acoustic signal, particularly for rare or domain-specific terms that are difficult to recognise from acoustics alone. Traditional ASR systems incorporate contextual biasing techniques ranging from shallow fusion with external language models ~\cite{zhao2019shallow,sriram2018coldfusion} to end-to-end contextual encoders that attend to bias phrases during decoding ~\cite{pundak2018clas,fox2022altspelling,tang2024guidedattn}. The ability to effectively exploit contextual information is thus an important capability for practical speech recognition systems.

Recent Audio Large Language Models (AudioLLMs) extend this paradigm by enabling transcription conditioned on free-form textual prompts. Models such as GPT-4o Transcribe~\cite{openai2024gpt4o}, the Gemini 3 family of models~\cite{gemini2025}, Sarvam Audio~\cite{sarvamaudio2026}, Gemma-3N~\cite{gemma3n2025}, Qwen3-Omni~\cite{xu2025qwen3omni}, and Voxtral~\cite{liu2025voxtral} accept audio alongside textual inputs that may include domain metadata, descriptions of the audio, or lists of relevant entities. In principle, this provides a flexible and unified interface for contextual ASR, allowing models to incorporate a wide range of contextual signals at inference time without specialised architectural mechanisms. However, it remains unclear whether these models genuinely utilise the provided prompts during transcription, or whether they rely primarily on parametric knowledge learned during pretraining. If a model correctly transcribes domain terminology regardless of the prompt content, the behaviour may reflect parametric memorisation rather than genuine contextual grounding.
\begin{table}[t]
  \caption{Comparison of contextual ASR benchmarks.}
  \label{tab:benchmark_comparison}
  \centering
  \footnotesize
  \setlength{\tabcolsep}{2pt}
  \begin{tabular}{@{}lcccl@{}}
    \toprule
    \textbf{Benchmark} & \textbf{Hours} & \textbf{Domains} & \textbf{Languages} & \textbf{Audio} \\
    \midrule
    IndicContextEval          & 56  & 23  & 8 (Indic)  & Natural \\
    ProfASR        & 8.6    & 4   & 1 (EN)     & Synthetic \\
    ContextASR     & 838   & 10+ & 2 (EN, ZH) & Synthetic \\
    Earnings-22    & 119   & 1   & 1 (EN)     & Natural \\
    \bottomrule
  \end{tabular}
\end{table}

Existing ASR benchmarks are not designed to answer this question (Table~\ref{tab:benchmark_comparison}). Large multilingual corpora such as IndicVoices \cite{javed2024indicvoices}, CommonVoice~\cite{ardila2020commonvoice}, and FLEURS~\cite{conneau2022fleurs} evaluate transcription under fixed prompting conditions with no variation of contextual inputs. Conversely, benchmarks designed for contextual ASR~\cite{wang2025contextasr,piskala2025profasr} focus primarily on English, often rely on synthetic speech, and typically test a single mode of context, such as, named entity lists. No existing evaluation systematically varies context types while holding other factors constant, making it unclear whether improvements arise from specific contextual signals or parametric memorisation.

To address this gap, we introduce \benchmark, a multilingual benchmark designed to evaluate contextual grounding and context utilisation in AudioLLMs. \benchmark~contains 56 hours of natural speech from 555 speakers across 8 Indian languages and 23 professional domains. We pair this dataset with a controlled prompt taxonomy consisting of seven levels (L0--L6), where each level introduces exactly one additional contextual signal, including domain metadata, natural-language audio descriptions, entity lists in English and native script, and adversarial prompts with incorrect entities. This controlled design enables attribution of performance changes to specific context types and allows systematic analysis of how models respond to contextual prompts. Our experiments reveal substantial differences in context utilisation behaviour: some models effectively exploit contextual information, others largely ignore it, and some respond unstably to prompt variations. This suggests that contextual grounding remains an open challenge for AudioLLMs despite their flexible prompting interfaces. All benchmark resources are publicly available at \url{https://github.com/AI4Bharat/IndicContextEval} to support future research.


\section{Related Work}

\noindent\textbf{Contextual Biasing in ASR.} Contextual information such as domain terminology or user-specific entities can significantly improve speech recognition. Early approaches incorporated such information through language model fusion, whether at decoding time~\cite{zhao2019shallow} or by jointly training the sequence model with a pretrained LM~\cite{sriram2018coldfusion}. Subsequent work introduced end-to-end contextual biasing mechanisms that encode contextual phrases and allow the decoder to attend to them during transcription~\cite{pundak2018clas}. Other work improves rare-word recognition via alternate spelling prediction~\cite{fox2022altspelling} and guided-attention losses that scale to large bias lists~\cite{tang2024guidedattn}.

In the context of modern ASR models, several mechanisms have been proposed for Whisper-style models, including contextual vocabulary injection~\cite{li2024cbwhisper}, dynamic vocabulary biasing~\cite{sudo2025owsmbiasing}, prompt-based domain vocabularies ~\cite{lall2024whispercontextbias}, supervised rare-word adaptation~\cite{jogi2025whisperrareword}, and pointer-based decoding with GPT-2~\cite{sun2023whispergpt2}. Contextual biasing has also been explored for LLM-based ASR through retrieval-based bias phrase selection~\cite{gong2025brasr}, reinforcement-learned hotword retrieval~\cite{kong2025hotwordrl}, and lightweight prompt-based biasing methods~\cite{ren2025lightpromptbias}. More broadly, prompting approaches enable domain adaptation, including zero-shot adaptation from domain descriptions~\cite{li2023llmpromptasr} and few-shot multilingual ASR via meta in-context learning~\cite{hsu2024smile}.
AudioLLMs further extend this paradigm by accepting textual prompts alongside audio. 
Recent systems, including commercial models~\cite{openai2024gpt4o, gemini2025, sarvamaudio2026} and open-weight approaches such as Gemma-3N~\cite{gemma3n2025}, Qwen3-Omni~\cite{xu2025qwen3omni}, and Voxtral~\cite{liu2025voxtral}, natively process both audio and text tokens. However,~\cite{yang2024prompts} show that Whisper’s prompt interface can behave unexpectedly, with corrupted prompts sometimes outperforming correct ones. 

\noindent\textbf{Benchmarks for Contextual ASR.} Existing benchmarks provide valuable test sets but leave critical questions unanswered. Large multilingual corpora such as 
~\cite{javed2024indicvoices, ardila2020commonvoice, conneau2022fleurs} prioritise language scale and speaker diversity but evaluate models under fixed prompting conditions with no variation of contextual inputs. Context-aware evaluation has largely focused on English, entity-rich domains: Earnings-22~\cite{rio2022earnings22} provides accented earnings calls, ContextASR-Bench~\cite{wang2025contextasr} spans multiple domains but relies on synthesised audio, and ProfASR-Bench~\cite{piskala2025profasr} targets professional speech with synthetic recordings. In contrast, \benchmark~combines multilingual natural speech with a controlled prompting framework that systematically varies contextual inputs, enabling analysis of how different context types influence transcription behaviour and allowing us to distinguish genuine contextual grounding from parametric memorisation.

\section{The \benchmark~ Benchmark}

\noindent\textbf{Design Goals:} The benchmark is designed to enable controlled evaluation of contextual grounding in AudioLLMs. To support this goal, the dataset satisfies 4 criteria. First, it contains natural speech across 8 Indian languages, covering diverse scripts and linguistic structures. Second, recordings span 23 professional domains, ensuring the presence of technical vocabulary and named entities that benefit from contextual information. Third, all recordings are paired with high-quality manual transcriptions produced by native speakers and verified through multi-stage quality control. Finally, each utterance is associated with structured contextual metadata and entity annotations, enabling systematic construction of contextual prompts. These design choices allow controlled analysis of how different contextual signals influence transcription behaviour in AudioLLMs.

\subsection{Domain Taxonomy}
To realise the domain diversity described above, the dataset spans 23 professional domains covering a wide range of technical, professional, and creative fields. These include areas such as \textit{Core Engineering}, \textit{Data Science}, \textit{Medical Sciences}, and \textit{Robotics \& Automation}; professional domains such as \textit{Forensics \& Legal Sciences}, \textit{Business}, and \textit{Defense \& Armed Forces}; as well as creative and cultural fields including \textit{Arts}, \textit{Film \& Media Production}, \textit{Culinary Arts \& Food Science}, and \textit{Linguistics}.

Each domain category further contains multiple sub-domains. For example, the \textit{Medical Sciences} category includes dental sciences, medical imaging, and clinical medicine. This hierarchical structure enables the systematic collection of domain-rich speech containing technical vocabulary and named entities across languages. This design ensures that many recordings contain terminology whose correct transcription can benefit from contextual prompting. A detailed list of all domain categories and their corresponding sub-domain descriptions is provided in the supplementary material.

\subsection{Dataset Creation}

The benchmark contains 55.93 hours of speech across 8 Indian languages: Hindi, Bengali, Telugu, Marathi, Gujarati, Malayalam, Odia, and Urdu, collected from 555 speakers with diverse backgrounds and recording conditions. Each language contributes at least 3 hours of speech, ranging from 3.37 hours for Urdu to 13.70 hours for Telugu, with an average of about 7 hours per language. Recordings were collected from students and professionals across diverse technical domains in two styles: \textit{read speech} and \textit{extempore speech}.

For extempore recordings, participants first selected three domains related to their expertise. For each selected domain, we prepared a set of domain-specific questions designed to cover diverse topics and encourage the use of varied domain vocabulary. Participants were then asked to speak about these questions, describing technical concepts, research topics, or professional experiences in a natural narrative or lecture-style manner. For read speech, domain-specific vocabulary lists were curated from textbooks and technical resources, and English sentences containing multiple domain terms were generated using Gemini 3 Pro. These sentences were translated into native languages using Sarvam-Translate \cite{sarvam_translate_2025}. All translations were reviewed and corrected by native-speaking language experts before recording.



\begin{figure}[t]
  \centering
  \begin{tikzpicture}
    \begin{axis}[
      width=0.9\columnwidth,
      height=5.5cm,
      xlabel={Context Level},
      ylabel={NEER (\%)},
      xtick={0,1,2,3,4,5,6},
      xticklabels={L0,L1,L2,L3,L4,L5,L6},
      legend style={font=\scriptsize, at={(0.5,-0.22)}, anchor=north, legend columns=4},
      grid=major,
      grid style={dashed, gray!30},
      ymin=0, ymax=45,
      tick label style={font=\footnotesize},
      label style={font=\footnotesize},
    ]
    \addplot[color=blue, mark=*, thick] coordinates
      {(0,36.67)(1,35.59)(2,34.82)(3,32.55)(4,34.50)(5,23.85)(6,34.55)};
    \addlegendentry{GPT-4o T}
    \addplot[color=red, mark=square*, thick] coordinates
      {(0,34.63)(1,25.85)(2,26.17)(3,24.69)(4,25.18)(5,17.39)(6,25.60)};
    \addlegendentry{Gemini 3F}
    \addplot[color=teal, mark=triangle*, thick] coordinates
      {(0,29.30)(1,25.93)(2,25.68)(3,25.05)(4,25.60)(5,21.69)(6,25.62)};
    \addlegendentry{Sarvam}
    \addplot[color=purple, mark=pentagon*, thick] coordinates
      {(0,40.12)(1,35.50)(2,36.24)(3,33.49)(4,34.61)(5,26.92)(6,36.25)};
    \addlegendentry{Gemma-3N}
    \end{axis}
  \end{tikzpicture}
  \caption{NEER (\%) across context levels. Native-script entities (L5) produce large drops for GPT-4o Transcribe, Gemini 3 Flash, and Gemma-3N, with a smaller effect on Sarvam Audio. L6 (adversarial) returns near L1 for all models. Gemini 3 Flash achieves the best NEER (17.39\% at L5).}
  \label{fig:context_neer}
\end{figure}
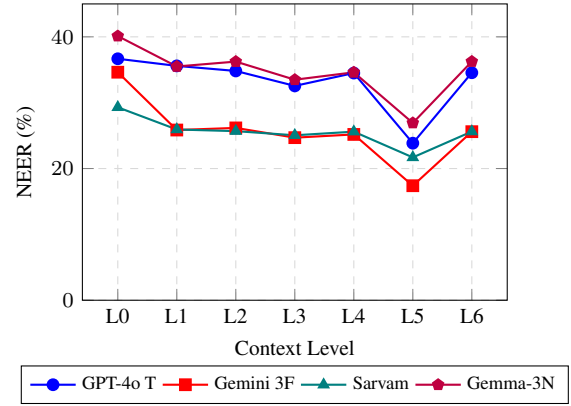

\subsection{Quality Control and Transcription}
All recordings were first manually verified to ensure recording quality and domain relevance before being sent for transcription. Quality control was performed by native speakers of the respective languages who were provided with domain terminology lists. 
Extempore recordings were accepted only if they demonstrated natural speech and sufficient use of domain-specific vocabulary. Read recordings were verified to ensure accurate reading and correct pronunciation of technical terms.

Reference transcriptions were produced by professional annotators who are native speakers of the target language, following guidelines similar to the IndicVoices ~\cite{javed2024indicvoices}. Speech was transcribed verbatim in the native script, preserving code-mixed segments. English named entities were transliterated into the native script and additionally provided in brackets in English. All transcripts were created from scratch without model assistance and underwent a multi-stage review process, with disagreements resolved by a senior annotator.

\subsection{Contextual Metadata and Entity Annotations}

Each utterance is associated with structured metadata used for contextual prompting and analysis:

\noindent \textbf{Domain label and description.} The domain category and a short topic description.
    
\noindent \textbf{Speech style.} Specifies whether speech is read or extempore.

\noindent \textbf{Region.} The speaker's geographic region.
    
\noindent \textbf{Named entities.} Domain-specific terminology curated by language experts and provided in English and the native script. The entity lists represent domain vocabulary and are used to construct entity-based contextual prompts during evaluation.

\noindent\textbf{Audio descriptions.} In addition to structured metadata, each audio segment includes a short natural-language description summarizing its topic and speaking style. These descriptions are generated using Gemini 3 Flash~\cite{gemini2025} from the available metadata. This enables comparison between structured metadata and natural-language context when evaluating AudioLLMs.

\subsection{Controlled Prompt Taxonomy (L0--L6)}
\label{sec:taxonomy}

To study contextual grounding in AudioLLMs, we design a controlled prompting framework with seven evaluation levels (L0--L6). Each level introduces exactly one additional contextual signal while keeping other factors constant, allowing performance changes to be attributed to specific context types. In all settings, models are required to produce output in the native script of the target language.

\noindent \textbf{L0 -- No context.} A bare transcription instruction with no language hint, measuring raw acoustic ASR and implicit language identification across Indic scripts.

\noindent \textbf{L1 -- Language only.} Target language is specified. This serves as the baseline for evaluating additional contextual signals.

\noindent \textbf{L2 -- Language + domain metadata.} A structured metadata block containing speech style, geographic region, and a one-sentence domain description. Tests whether recording context improves transcription.

\noindent \textbf{L3 -- Language + audio description.} A short natural-language description of the audio's topic and discussion type generated from the same metadata as L2. Tests whether natural-language context is more effective than structured metadata.

\noindent \textbf{L4 -- Language + entities (English).} A list of 20--30 domain entities in English, randomly sampled from the domain vocabulary, is provided while output remains in the native script, testing cross-lingual entity biasing. The entities provided may or may not appear in the audio, but provide domain context.

\noindent \textbf{L5 -- Language + entities (native script).} The same entity list as L4, but written in native script, aligning prompt and output. The L5--L4 difference measures the \textit{script mismatch cost}.

\noindent \textbf{L6 -- Wrong entities (adversarial).} A list of 20--30 entities from an unrelated domain is provided in the native script (e.g., medical entities for robotics audio). If performance drops relative to L1, the model is influenced by entity prompts; if not, it ignores them.

\section{Experimental Setup}

\subsection{Models evaluated}

We evaluate 5 models on our benchmark, selecting leading proprietary and open-weight AudioLLMs that claim support for all 8 languages in our dataset, alongside a strong standalone ASR baseline.

\noindent\textbf{1. Standalone ASR baseline} (evaluated at L1 only): We evaluate \emph{IndicConformer}~\cite{indicconformer2023}, a 600M-parameter multilingual Conformer trained on 22 Indian languages. Since it requires the target language as input, it cannot operate at L0 and is evaluated at L1 as a competitive non-LLM reference.


\noindent\textbf{2. AudioLLMs} (evaluated at all seven levels, L0--L6): We evaluate \emph{GPT-4o Transcribe}~\cite{openai2024gpt4o}, \emph{Gemini 3 Flash}~\cite{gemini2025}, \emph{Sarvam Audio}~\cite{sarvamaudio2026} for commercial models and select \emph{Gemma-3N}~\cite{gemma3n2025} (8B-E4B) for open weight models. We do not include models such as \emph{Qwen3-Omni}~\cite{xu2025qwen3omni} and \emph{Voxtral}~\cite{liu2025voxtral} because their official documentation does not claim support for all eight Indian languages evaluated in this work.

\subsection{Prompting protocol}

Every prompt follows a fixed two-part structure: a context block followed by a transcription instruction. The context block is empty at L0, contains only language at L1, and adds one additional context type at L2--L6. The transcription instruction is identical across all models and context levels: output must be in the native script, numbers must be written as spoken words, hesitations must be ignored, and English words must be transliterated into the native script.

\subsection{Evaluation metrics}

\noindent\textbf{Word Error Rate (WER):} Edit distance normalised by reference length. Text normalisation uses the Indic NLP Library~\cite{kakwani2020indicnlpsuite} for Indic languages.

\noindent\textbf{Named Entity Error Rate (NEER):} Fraction of reference named entities absent or incorrectly transcribed. NEER is the primary metric for entity biasing (L4--L6).

\begin{table}[t]
  \caption{WER and NEER (\%) at L1 (language specified).}
  \label{tab:baseline}
  \centering
  \footnotesize
  \setlength{\tabcolsep}{4pt}
  \begin{tabular}{@{}llrr@{}}
    \toprule
    \textbf{Type} & \textbf{Model} & \textbf{WER} & \textbf{NEER}\\
    \midrule
    ASR
      & IndicConformer & 18.81 & 29.58 \\
    \midrule
    \multirow{4}{*}{AudioLLM}
      & Sarvam Audio      & 16.86 & 25.93 \\
      & Gemini 3 Flash    & 18.90 & 25.85 \\
      & GPT-4o Transcribe & 28.61 & 35.59 \\
      & Gemma-3N          & 38.73 & 35.50 \\
    \bottomrule
  \end{tabular}
\end{table}

\section{Results}
\label{sec:results}

\subsection{Baseline performance}

Table~\ref{tab:baseline} reports the average WER at L1 (language prompt) for all models. Sarvam Audio achieves the lowest WER, followed by IndicConformer and Gemini 3 Flash, while GPT-4o Transcribe and Gemma-3N show substantially higher error rates.


\subsection{Contextual sensitivity across models}

Table~\ref{tab:wer_by_level} reports WER across all 7 context levels. Figure~\ref{fig:context_neer} shows the corresponding NEER trajectories. Three key patterns emerge:

\begin{table}[t]
  \caption{WER (\%) by prompt level.}
  \label{tab:wer_by_level}
  \centering
  \footnotesize
  \setlength{\tabcolsep}{3pt}
  \begin{tabular}{@{}l rrrrrrr@{}}
    \toprule
    \textbf{Model} & \textbf{L0} & \textbf{L1} & \textbf{L2} & \textbf{L3} & \textbf{L4} & \textbf{L5} & \textbf{L6} \\
    \midrule
    GPT-4o T   & 29.83 & 28.61 & 28.37 & 26.08 & 27.97 & \textbf{26.04} & 28.47 \\
    Gemini 3F  & 24.30 & 18.90 & 19.28 & 18.39 & 19.88 & \textbf{17.46} & 19.67 \\
    Sarvam     & 20.39 & 16.86 & 16.78 & 16.43 & 16.80 & \textbf{15.70} & 16.69 \\
    Gemma-3N   & 51.21 & \textbf{38.73} & 52.20 & 40.22 & 46.37 & 43.11 & 47.95 \\
    \bottomrule
  \end{tabular}
\end{table}

\begin{table}[h]
    \centering
    \footnotesize
    \setlength{\tabcolsep}{2.5pt}
    \newcommand{\cell}[2]{\cellcolor{red!#1}#2}
    \caption{Per-language WER (\%) at L5. Darker = higher error.}
    \begin{tabular}{@{}l cccccccc@{}}
      \toprule
      & \textbf{Hi} & \textbf{Bn} & \textbf{Te} & \textbf{Mr} & \textbf{Gu} & \textbf{Ml} & \textbf{Or} & \textbf{Ur} \\
      \midrule
      GPT-4o T   & \cell{14}{17.5} & \cell{14}{18.5} & \cell{24}{30.9} & \cell{20}{24.3} & \cell{24}{31.0} & \cell{34}{42.6} & \cell{26}{31.9} & \cell{16}{19.6} \\
      Gemini 3F  & \cell{12}{14.3} & \cell{10}{13.3} & \cell{18}{22.5} & \cell{12}{14.5} & \cell{10}{12.0} & \cell{24}{29.7} & \cell{14}{18.8} & \cell{14}{18.6} \\
      Sarvam     & \cell{10}{12.4} & \cell{10}{12.7} & \cell{14}{18.7} & \cell{10}{13.2} & \cell{10}{11.4} & \cell{24}{30.8} & \cell{12}{15.5} & \cell{16}{20.0} \\
      Gemma-3N   & \cell{26}{33.2} & \cell{22}{28.5} & \cell{34}{42.7} & \cell{30}{37.5} & \cell{46}{57.4} & \cell{56}{70.2} & \cell{46}{58.4} & \cell{34}{43.1} \\
      \bottomrule
    \end{tabular}
    \label{tab:heatmap_lang}
\end{table}

\noindent \textbf{Language identification significantly affects performance.}The transition from L0 (no context) to L1 (language specified) reveals a substantial language-identification tax. Gemma-3N improves by 12.48 WER points, Gemini 3 Flash by 5.40, and Sarvam Audio by 3.53, while GPT-4o Transcribe improves by just 1.22 points. Without a language hint, models must infer the correct script from acoustics alone, leading to errors because the transcription is produced in the wrong script.

\noindent \textbf{The form of context matters as much as its content.} Natural-language audio descriptions (L3) consistently outperform structured metadata prompts (L2). GPT-4o Transcribe improves by 2.53 WER points at L3 compared to only 0.24 at L2, while Gemini 3 Flash gains 0.51 WER at L3 but slightly regresses at L2 (+0.38). Gemma-3N is particularly sensitive to prompt format: structured metadata severely degrades performance (+13.47 WER), whereas the equivalent natural-language description produces only a minor degradation (+1.49 WER). Thus, identical contextual information can yield markedly different outcomes depending on how it is expressed.

\noindent \textbf{Native-script entity biasing provides the strongest gains.} Supplying entity lists as domain context in the target language script (L5) produces the largest NEER improvements across models (Figure~\ref{fig:context_neer}). GPT-4o Transcribe improves by 11.7\% 
, Gemini 3 Flash by 8.5\% 
, Gemma-3N by 8.6\% 
, and Sarvam Audio by 4.2\% 
. The gap between English-script entities (L4) and native-script entities (L5) reaches up to 11 points, confirming a substantial script-mismatch cost. Table~\ref{tab:heatmap_lang} breaks down L5 WER by language, revealing large cross-lingual variation: Malayalam is the hardest language for most models.

\subsection{Adversarial control (L6)}

L6 serves as a negative control: if models genuinely rely on entity prompts, providing entities from an incorrect domain should degrade performance. Table~\ref{tab:adversarial} compares L6 against the L1 baseline and reveals four distinct behaviours. \textbf{GPT-4o Transcribe} is adversarially robust (L6 $\approx$ L1) while still benefiting from correct entities at L5 ($-$2.57 WER), suggesting that the model cross-validates entity hints against acoustic evidence. \textbf{Gemini 3 Flash} shows moderate sensitivity (+0.77 WER), indicating that its L5 improvements reflect genuine contextual use. In contrast, \textbf{Gemma-3N} is heavily degraded by adversarial entities (+9.22 WER), indicating blind reliance on entity prompts. Finally, \textbf{Sarvam Audio} remains largely unaffected (L6 $\approx$ L1), consistent with its minimal sensitivity to textual context observed across earlier levels.

\begin{table}[!]
  \caption{Adversarial control: L6 (wrong entities) vs.\ L1.}
  \label{tab:adversarial}
  \centering
  \footnotesize
  \setlength{\tabcolsep}{3pt}
  \begin{tabular}{@{}l rr r rr r@{}}
    \toprule
    & \multicolumn{3}{c}{\textbf{WER}} & \multicolumn{3}{c}{\textbf{NEER}} \\
    \cmidrule(lr){2-4} \cmidrule(lr){5-7}
    \textbf{Model} & \textbf{L1} & \textbf{L6} & \textbf{$\Delta$} & \textbf{L1} & \textbf{L6} & \textbf{$\Delta$} \\
    \midrule
    GPT-4o T   & 28.61 & 28.47 & $-$0.14 & 35.59 & 34.55 & $-$1.04 \\
    Gemini 3F  & 18.90 & 19.67 & +0.77   & 25.85 & 25.60 & $-$0.25 \\
    Sarvam     & 16.86 & 16.69 & $-$0.17 & 25.93 & 25.62 & $-$0.31 \\
    Gemma-3N   & 38.73 & 47.95 & +9.22   & 35.50 & 36.25 & +0.75 \\
    \bottomrule
  \end{tabular}
\end{table}

\subsection{Discussion}

\noindent\textbf{Balanced utilisation (GPT-4o Transcribe):} This model benefits from correct contextual signals ($-\Delta$2.57 WER at L5) while remaining robust to incorrect prompts (L6 $\approx$ L1). Although its baseline WER is higher than other production models, GPT-4o demonstrates the most reliable contextual reasoning, selectively exploiting useful prompts while ignoring adversarial ones.

\noindent\textbf{Sensitive utilisation (Gemini 3 Flash):} Gemini 3 Flash shows strong improvements when relevant context is provided, achieving the best entity accuracy overall (17.39\% NEER at L5) and a substantial WER reduction ($-$1.44 from L1 to L5). Most samples benefit from contextual prompts, indicating consistent integration of contextual information during decoding.

\noindent\textbf{Unstable utilisation (Gemma-3N):} Gemma-3N exhibits improvements in entity recognition (35.5\%$\rightarrow$26.9\% NEER at L5) but instability in overall transcription quality. WER increases by 4.38 points from L1 to L5, and 13.2\% of L5 samples exhibit severe hallucinations or repetitions. The model identifies contextual entities but often corrupts the surrounding transcript.

\noindent\textbf{Context-blind behaviour (Sarvam Audio):} Sarvam Audio achieves the lowest baseline WER among AudioLLMs (16.86\% at L1) and improves slightly to 15.70\% at L5, with minimal variation across context levels (1.16 WER gain). This suggests transcription is largely driven by the acoustic encoder, with limited influence from textual prompts.

\noindent\textbf{Context narrows the ASR gap:} At L5, even Gemini 3 Flash surpasses the standalone ASR baseline (IndicConformer, 18.81\% WER) with 17.46\% WER. For entity recognition, all four AudioLLMs at L5 beat IndicConformer's NEER of 29.58\%, with Gemini 3 Flash achieving the lowest at 17.39\%.


\section{Conclusion}

We introduced \benchmark, a multilingual benchmark and controlled prompt taxonomy for evaluating context utilisation in AudioLLMs. \benchmark~spans 55.93 hours of natural speech across eight Indian languages and 23 domains, enabling systematic analysis of how different contextual signals affect transcription. Our experiments show that models vary widely in how they use context. Native-script entity biasing yields the largest gains, and natural-language descriptions consistently outperform structured metadata prompts. Adversarial prompts further reveal divergent behaviours: some models selectively use context, while others ignore it or rely on it blindly. These findings indicate that contextual grounding remains an open challenge for AudioLLMs.
\ifcameraready
\section{Acknowledgments}
We gratefully acknowledge the support of EkStep Foundation and Nilekani Philanthropies, whose generous funding made this work possible by supporting the team, resources, and cloud infrastructure required for the project. We sincerely thank the NPTEL team at IIT Madras for their invaluable assistance in reaching and engaging participants for the data collection effort. We extend our heartfelt appreciation to all participants who generously contributed their time and voices to this study; this work would not have been possible without their support. We further thank the AI4Bharat linguistics team, project management team, data leads, and all members who contributed to the planning, execution, and successful completion of this project.

\fi

\section{Generative AI Use Disclosure}
Generative AI tools were used solely for language polishing and editing during the preparation of this manuscript. These tools assisted with improving clarity, grammar, and conciseness of the writing. No generative AI system was used to generate experimental results, analyses, figures, or scientific conclusions. All technical content, experiments, and interpretations were developed and verified by the authors.
\bibliographystyle{IEEEtran}
\bibliography{mybib}

@inproceedings{zhao2019shallow,
  title={Shallow-fusion end-to-end contextual biasing},
  author={Zhao, Ding and Sainath, Tara N. and Rybach, David and Rondon, Pedro and Bhatia, Deepak and Li, Bo and Pang, Ruoming},
  booktitle={Proc. Interspeech},
  pages={1418--1422},
  year={2019}
}

@inproceedings{pundak2018clas,
  title={Deep context: End-to-end contextual speech recognition},
  author={Pundak, Golan and Sainath, Tara N. and Prabhavalkar, Rohit and Kannan, Anjuli and Zhao, Ding},
  booktitle={Proc. IEEE SLT},
  pages={418--425},
  year={2018}
}

@misc{openai2024gpt4o,
  author={OpenAI},
  title={GPT-4o Transcribe model},
  year={2024},
  url={https://developers.openai.com/api/docs/models/gpt-4o-transcribe}
}

@misc{gemini2025,
  author={{Google DeepMind}},
  title={Gemini 3},
  year={2025},
  url={https://deepmind.google/models/gemini/}
}

@misc{sarvamaudio2026,
  author={{Sarvam AI}},
  title={Sarvam Audio: Speech recognition beyond transcription},
  year={2026},
  url={https://www.sarvam.ai/blogs/sarvam-audio}
}

@misc{gemma3n2025,
  author={Google},
  title={Gemma 3n: Powerful, efficient, mobile-first AI},
  year={2025},
  url={https://developers.googleblog.com/en/introducing-gemma-3n}
}

@article{xu2025qwen3omni,
  title={Qwen3-Omni technical report},
  author={Xu, J. and others},
  journal={arXiv preprint arXiv:2509.17765},
  year={2025}
}

@article{liu2025voxtral,
  title={Voxtral},
  author={Liu, A. H. and others},
  journal={arXiv preprint arXiv:2507.13264},
  year={2025}
}

@inproceedings{javed2024indicvoices,
  title={IndicVoices: Towards building an inclusive multilingual speech dataset for Indian languages},
  author={Javed, T. and others},
  booktitle={Findings of ACL},
  pages={10740--10782},
  year={2024}
}

@inproceedings{ardila2020commonvoice,
  title={Common Voice: A massively-multilingual speech corpus},
  author={Ardila, R. and others},
  booktitle={Proc. LREC},
  pages={4218--4222},
  year={2020}
}

@inproceedings{conneau2022fleurs,
    author={Conneau, Alexis and Ma, Min and Khanuja, Simran and Zhang, Yu and Axelrod, Vera and Dalmia, Siddharth and Riesa, Jason and Rivera, Clara and Bapna, Ankur},
  booktitle={2022 IEEE Spoken Language Technology Workshop (SLT)}, 
  title={FLEURS: FEW-Shot Learning Evaluation of Universal Representations of Speech}, 
  year={2023},
  volume={},
  number={},
  pages={798-805},
  doi={10.1109/SLT54892.2023.10023141}
  }

@article{wang2025contextasr,
  title={ContextASR-Bench: A massive contextual speech recognition benchmark},
  author={Wang, H. and Ma, L. and Guo, D. and Wang, X. and Xie, L. and Xu, J. and Lin, J.},
  journal={arXiv preprint arXiv:2507.05727},
  year={2025}
}

@article{piskala2025profasr,
  title={ProfASR-Bench: A benchmark for context-conditioned ASR in high-stakes professional speech},
  author={Piskala, D. B.},
  journal={arXiv preprint arXiv:2512.23686},
  year={2025}
}

@inproceedings{sriram2018coldfusion,
  title={Cold Fusion: Training Seq2Seq models together with language models},
  author={Sriram, Anirudh and Jun, H. and Satheesh, S. and Coates, A.},
  booktitle={Proc. Interspeech},
  pages={387--391},
  year={2018}
}

@article{fox2022altspelling,
  title={Improving contextual recognition of rare words with an alternate spelling prediction model},
  author={Drexler Fox, Jennifer and Delworth, Natalie},
  journal={arXiv preprint arXiv:2209.01250},
  year={2022}
}

@inproceedings{tang2024guidedattn,
  author={Tang, Jiyang and Kim, Kwangyoun and Shon, Suwon and Wu, Felix and Sridhar, Prashant},
  booktitle={ICASSP 2024 - 2024 IEEE International Conference on Acoustics, Speech and Signal Processing (ICASSP)}, 
  title={Improving ASR Contextual Biasing with Guided Attention}, 
  year={2024},
  volume={},
  number={},
  pages={12096-12100},
  doi={10.1109/ICASSP48485.2024.10447438}}

@inproceedings{li2024cbwhisper,
  title={CB-Whisper: Contextual biasing Whisper using open-vocabulary keyword-spotting},
  author={Li, Y. and others},
  booktitle={Proc. LREC-COLING},
  pages={2941--2946},
  year={2024}
}

@article{sudo2025owsmbiasing,
  title={OWSM-Biasing: Contextualizing open Whisper-style speech models for ASR with dynamic vocabulary},
  author={Sudo, Y. and Fujita, Y. and Kojima, A. and Mizumoto, T. and Liu, L.},
  journal={arXiv preprint arXiv:2506.09448},
  year={2025}
}

@article{lall2024whispercontextbias,
  title={Contextual biasing to improve domain-specific custom vocabulary audio transcription without explicit fine-tuning of Whisper model},
  author={Lall, V. and Liu, Y.},
  journal={arXiv preprint arXiv:2410.18363},
  year={2024}
}

@article{jogi2025whisperrareword,
  title={Improving rare-word recognition of Whisper in zero-shot settings},
  author={Jogi, Y. and Aggarwal, V. and Nair, S. S. and Verma, Y. and Kubba, A.},
  journal={arXiv preprint arXiv:2502.11572},
  year={2025}
}

@article{sun2023whispergpt2,
  title={Can contextual biasing remain effective with Whisper and GPT-2?},
  author={Sun, G. and Zheng, X. and Zhang, C. and Woodland, P. C.},
  journal={arXiv preprint arXiv:2306.01942},
  year={2023}
}

@article{gong2025brasr,
  title={BR-ASR: Efficient and scalable bias retrieval framework for contextual biasing ASR in speech LLM},
  author={Gong, X. and Lv, A. and Wang, Z. and Zhu, H. and Qian, Y.},
  journal={arXiv preprint arXiv:2505.19179},
  year={2025}
}

@article{kong2025hotwordrl,
  title={Contextual biasing for LLM-based ASR with hotword retrieval and reinforcement learning},
  author={Kong, Y. and Hou, J. and Tang, J. and Zhu, B. and Zhang, J. and Xue, S.},
  journal={arXiv preprint arXiv:2512.21828},
  year={2025}
}

@article{ren2025lightpromptbias,
  title={Lightweight prompt biasing for contextualized end-to-end ASR systems},
  author={Ren, B. and Shi, Y. and Li, J.},
  journal={arXiv preprint arXiv:2506.06252},
  year={2025}
}

@article{li2023llmpromptasr,
  title={Prompting large language models for zero-shot domain adaptation in speech recognition},
  author={Li, Y. and Wu, Y. and Li, J. and Liu, S.},
  journal={arXiv preprint arXiv:2306.16007},
  year={2023}
}

@article{hsu2024smile,
  title={SMILE: Speech meta in-context learning for low-resource language automatic speech recognition},
  author={Hsu, M. H. and Lee, H. Y.},
  journal={arXiv preprint arXiv:2409.10429},
  year={2024}
}

@inproceedings{yang2024prompts,
  author={Yang, Chih-Kai and Huang, Kuan-Po and Lee, Hung-Yi},
  booktitle={2024 IEEE Spoken Language Technology Workshop (SLT)}, 
  title={Do Prompts Really Prompt? Exploring the Prompt Understanding Capability of Whisper}, 
  year={2024},
  volume={},
  number={},
  pages={1-8},
  doi={10.1109/SLT61566.2024.10832185}
  }

@inproceedings{rio2022earnings22,
  title={Earnings-22: A practical benchmark for accents in the wild},
  author={Del Rio, M. and Ha, P. and McNamara, Q. and Miller, C. and Chandra, S.},
  booktitle={Proc. Interspeech},
  year={2022}
}

@misc{indicconformer2023,
  title={IndicConformer: Multilingual ASR model for 22 Indian languages},
  author={{AI4Bharat}},
  year={2024},
  url={https://huggingface.co/ai4bharat/indic-conformer-600m-multilingual}
}

@inproceedings{kakwani2020indicnlpsuite,
  title={IndicNLPSuite: Monolingual corpora, evaluation benchmarks and pre-trained multilingual language models for Indian languages},
  author={Kakwani, Divyanshu and Kunchukuttan, Anoop and Golla, Satish and others},
  booktitle={Findings of EMNLP},
  pages={4948--4961},
  year={2020}
}

@misc{sarvam_translate_2025,
  title        = {Sarvam-Translate},
  author       = {{Sarvam AI}},
  year         = {2025},
  howpublished = {\url{https://huggingface.co/sarvamai/sarvam-translate}},
  note         = {Hugging Face model repository},
}

\end{document}